Editorial v5 4/7/2010

# Dwarf-Galaxy Cosmology Editorial

Regina Schulte-Ladbeck[1], Ulrich Hopp[2], Elias Brinks[3], and Andrey Kravtsov[4],

[1]Department of Physics and Astronomy, University of Pittsburgh, Pittsburgh, PA 15260, USA
[2]Universitäts-Sternwarte, Ludwig-Maximilians-Universität München, 81679 München, Germany; Max-Planck Institut für Extraterrestrische Physik, 85748 Garching, Germany
[3]Centre for Astrophysics Research, University of Hertfordshire, Hatfield AL10 9AB, UK
[4]Kavli Institute for Cosmological Physics, The Department of Astronomy and Astrophysics, The University of Chicago, Chicago, IL 60637, USA

Dwarf galaxies provide opportunities for drawing inferences about the processes in the early universe by observing our "cosmological backyard" – the Local Group and its vicinity. This special issue of the open-access journal Advances in Astronomy is a snapshot of the current state of the art of dwarf-galaxy cosmology. The issue contains fourteen review papers, and one original research article. All papers were peer-reviewed by a minimum of two referees.

Dwarf galaxies continue to challenge our cosmological models and models of galaxy formation. Two well know problems are the "missing satellites problem" and the "cores versus cusps problem."

The dilemma posed by the fact that observations seem to indicate an approximately constant dark matter density in the inner parts of galaxies, while cosmological simulations prefer a steep, power-law-like behavior is assessed by W.J.G. de Blok in the paper "*The Core-Cusp Problem*." Andrey Kravtsov reviews the quandary that predicted subhalos outnumber observed dwarf galaxies of the Local Group. In his paper "*Dark Matter Substructure and Dwarf Galactic Satellites*," he emphasizes insights that have been gained from cosmological simulations and their tension with observational data.

The observer's viewpoint of the missing satellites problem is provided by Beth Willman in the review "*In Pursuit of the Least Luminous Galaxies*." Her paper focuses on the progress made with the help of the Sloan Digital Sky Survey data, and gives perspectives on the potential for contributions to this research by future surveys and new telescopes. Her paper is complemented by Helmut Jerjen's review "*Dwarf Cosmology with the Stromlo Missing Satellites Survey*." He makes the case that the southern hemisphere has been largely unexplored to faint magnitude levels with modern digital imaging data, and discusses the prospects for a complete census of the Milky Way satellite population. An approach to finding missing dwarf galaxies that is independent of their luminous stellar content is through detecting their gravitational lensing signal. Eric Zackrisson and Teresa Riehm examine the many ways through which lensing by subhalos can manifest itself in their paper titled "*Gravitational Lensing as a Probe of Cold Dark Matter Subhalos*."

Early cosmological simulations also predict large numbers of dwarf galaxies in voids, which, since this is not confirmed by observations either, is another face of the missing dwarfs issue. It is currently hoped that a better understanding of the physics that controls star formation might resolve the discrepancies between the simulated and observed dwarf galaxies in both environments. Work has therefore focused on how gas is accreted and retained in dwarf-galaxy-sized halos to form stars. In "*The First Galaxies and the Likely Discovery of their Fossils in the Local Group*," Massimo Ricotti reviews simulations of the formation and fate of pre-ionization dwarfs. Matthias Hoeft and Stefan Gottlöber describe what progress has been made with high-resolution simulations to tackle the issue of "*Dwarf Galaxies in Voids: Dark Matter Halos and Gas Cooling*." Kentaro Nagamine adds his views on how to include star formation processes in cosmological simulations in "*Star Formation History of Dwarf Galaxies in Cosmological Hydrodynamic Simulations*."

These papers are accompanied by a review of the current knowledge of the star-formation histories of dwarf galaxies that has been gained from observations. Michele Cignoni and Monica Tosi describe experimental approaches and results in "*Star Formation Histories of Dwarf Galaxies from the Colour-Magnitude Diagrams of Their Resolved Stellar Populations*."

Several papers address how dynamical interactions influence the evolution of dwarf galaxies. In "*Environmental Mechanisms Shaping the Nature of Dwarf Galaxies: The View of Computer Simulations*," Lucio Mayer describes a mechanism for transforming disky dwarfs into dwarf spheroidals, and discusses implications for the substructure problem. Laura Virginia Sales, Amina Helmi, and Giuseppina Battaglia consider "*The Effect of Tidal Stripping on Composite Stellar Populations in Dwarf Spheroidal Galaxies*." Their dwarf galaxy model follows the evolution of two kinematically and spatially segregated stellar components. They ask how these can be distinguished after having interacted with the potential of a massive galaxy. Frederic Bournaud, in "*Tidal Dwarf Galaxies and Missing Baryons*," discusses a tension that exists between model predictions that tidal dwarfs should not contain a significant mass fraction from the dark matter halos of their progenitor spiral galaxies, and existing observations which do suggest the presence of an unseen component.

Dwarf spheroidal galaxies are considered to provide good astronomical sites for the study of the nature of dark matter. In "*Kinematics of Milky Way Satellites: Mass Estimates, Rotation Limits, and Proper Motions,*" Louis Strigari reviews the evidence for the presence of dark matter in dwarf spheroidal companions of the Milky Way. Michael Kuhlen portrays prospects for the detection of the hypothetical dark matter particle. His paper is titled "*The Dark Matter Annihilation Signal from Dwarf Galaxies and Subhalos*." Arthur Kosowsky makes the case that dwarf spheroidals are not only important for testing the dark matter hypothesis, but also the competing, modified Newtonian gravity hypothesis. His paper critically analyzes "*Dwarf Galaxies, MOND, and Relativistic Gravitation*."

Editorial v5 4/7/2010

Arthur Kosowsky aptly summarizes what we consider the spirit of this special issue: "Until the predominant picture of dark matter cosmology can explain all of the observations, other competing ideas should be pursued, either as a way of sharpening the case for dark matter cosmology, or, perhaps, uncovering an eventual replacement. […] We should not allow the successes of our leading theories to blind us to other possibilities."

There is much opportunity for theoretical and observational innovation in the area of dwarf-galaxy cosmology. We asked the authors of our review papers to write them in the style of a tutorial and at a level suitable for beginning graduate students. We hope our special issue will reach a wide audience of graduate students and beginning researchers, in particular since all of the papers, under the open-access model, are accessible free of charge to anyone with a computer and an internet connection.

We would like to thank the members of the editorial board of Advances in Astronomy, the authors and referees of articles submitted to the special issue, and the staff of Hindawi publisher. Without their support and hard work, this special issue would not have come into being.

We hope you will enjoy reading *Dwarf-Galaxy Cosmology*.